\documentclass[times,twocolumn,trackchanges]{aastex62}
\usepackage{graphicx}
\usepackage{amsmath}
\usepackage{color}

\turnoffedit

\newcommand{\mathsc}[1]{{\text{\normalfont\scshape#1}}} 
\newcommand\ho{\ifmmode {\rm H\hspace{.1em}\mathsc{i}} \else \mbox{\rm H\,\scshape{i}} \fi}
\def\aG{\ifmmode {\alpha G}\else $\alpha G$ \fi}
\def\sg{\ifmmode \sigma_{\rm g} \else $\sigma_{\rm g}$ \fi}

\newcommand{\hi}{\mbox{\rm H$\,$\scshape{i}}}

\newcommand{\Kkmpers}{\mbox{K~km~s$^{-1}$}}

\newcommand{\MJypersr}{\mbox{MJy~sr$^{-1}$}}

\newcommand{\Msunperpc}{\mbox{\rm M$_{\odot}$ pc$^{-2}$}}
\newcommand{\Msun}{\mbox{$\rm M_{\odot}$}}

\newcommand{\Msunperyrperkpc}{\mbox{\rm M$_{\odot}$ yr$^{-1}$ kpc$^{-2}$}}

\newcommand{\Msunperyr}{\mbox{\rm M$_{\odot}$ yr$^{-1}$}}

\shorttitle{Metallicity Dependence of \hi\ Shielding Layers}
\shortauthors{Schruba et al.}

\accepted{May 11, 2018}
\submitjournal{The Astrophysical Journal}

\begin{document}
\title{The Metallicity Dependence of the \hi\ Shielding Layers in Nearby Galaxies}

\author{Andreas Schruba}
\affiliation{Max-Planck-Institut f\"ur extraterrestrische Physik, Giessenbachstra{\ss}e 1, D-85748 Garching, Germany}

\author{Shmuel Bialy}
\affiliation{Raymond and Beverly Sackler School of Physics \& Astronomy, Tel Aviv University, Ramat Aviv 69978, Israel}

\author{Amiel Sternberg}
\affiliation{Raymond and Beverly Sackler School of Physics \& Astronomy, Tel Aviv University, Ramat Aviv 69978, Israel}

\correspondingauthor{Andreas Schruba}
\email{schruba@mpe.mpg.de}

\defcitealias{Sternberg14}{S14}

\begin{abstract}
We investigate the metallicity dependence of \hi\ surface densities in star-forming regions along many lines of sight within 70 nearby galaxies, probing kpc to $50$~pc scales. We employ \hi, SFR, stellar mass, and metallicity (gradient) measurements from the literature, spanning a wide range ($5$~dex) in stellar and gas mass and ($1.6$~dex) in metallicity. \edit1{We consider metallicities as observed, or rescaled} to match the mass--metallicity relation determined for SDSS galaxies. At intermediate to high metallicities ($0.3\,{-}\,2$ times solar), we find that the \hi\ surface densities saturate at sufficiently large total gas surface density. The maximal \hi\ columns vary approximately inversely with metallicity, and show little variation with spatial resolution, galactocentric radius, or among galaxies. In the central parts of massive spiral galaxies the \hi\ gas is depressed by factors ${\sim}\,2$. The observed behavior is naturally reproduced by metallicity dependent shielding theories for the \hi-to-H$_2$ transitions in star-forming galaxies. We show that the inverse scaling of the maximal \hi\ columns with metallicity suggests that the area filling fraction of atomic-molecular complexes in galaxies is of order unity, and weakly dependent on metallicity.
\end{abstract}

\section{Introduction}
\label{sec:intro}

The atomic-to-molecular hydrogen (\hi-to-H$_2$) transition is of fundamental importance for the evolution of the interstellar medium (ISM) and for star formation in galaxies. Stars form in cold dense clouds composed of molecular gas \citep[see reviews by][]{McKee07, Kennicutt12, Tan14}. Observations have revealed the close small-scale connection between the dense gas mass and the star formation rate (SFR) in the Milky Way \citep{Evans14, Stephens16, Shimajiri17} and in local galaxies \citep{Gao04, Usero15, Bigiel16}. On large kpc-scales the SFR appears to correlate with the bulk molecular gas mass traced by CO emission \citep{Bigiel11, Schruba11, Leroy13b}. However, the ISM of (most) galaxies in the local universe is dominated by their \hi\ reservoirs \citep{Saintonge11a}. The SFR per unit \hi\ mass varies by orders of magnitude on galactic scales \citep{Huang12, Wang17} and is entirely unrelated to star formation in the H$_2$-dominated parts of spiral galaxies \citep{Bigiel11, Schruba11}. Therefore, understanding how the \hi-to-H$_2$ transition is related to local and global galactic properties, e.g., mass, morphology, and metallicity, is critical for predicting the dense gas fractions and SFRs of galaxies.

Observational studies of the ratio of molecular to atomic hydrogen mass surface densities, $R_{\rm mol} \equiv \Sigma_{\rm H_2} / \Sigma_\ho$, have focused on either the hydrostatic midplane pressure and thus on midplane volume density \citep{Wong02, Blitz06, Leroy08}, or on total gas surface density and its variation with metallicity \citep{Fumagalli10, Schruba11, Wong13}. In theoretical models of the \hi-to-H$_2$ transition, the H$_2$ formation on dust grains as well as dust attenuation of dissociating ultraviolet (UV) radiation scale with dust-to-gas ratio, and hence metallicity \citep{Krumholz09a, Krumholz13, Sternberg14, Bialy16}. Additional dependencies on volume density and radiation field strength also enter in the H$_2$ formation and dissociation processes.

Recently, highly resolved observations of \hi\ and infrared emission by dust of molecular clouds in the Milky Way and the Magellanic Clouds found good agreement between observations and models based on dust-shielding \citep{Wong09, Bolatto11, LeeMY12, LeeC15, RomanDuval14, Bialy15b, Bialy17a}, but for a narrow range of metallicities. So far only a single study by \citet{Fumagalli10} has focused on low metallicity galaxies. For a small sample of blue compact (star-bursting) dwarfs (BCDs), they argue that on scales ${<}\,100$~pc, the chemical composition of the ISM is mostly sensitive to the gas surface density and metallicity rather than hydrostatic pressure. However, their study is also significantly limited by the small sample, heterogeneous data sets, and assumptions about the gas geometry within the telescope beam. 

In this paper we study the metallicity dependence of the maximal \hi\ surface densities at high spatial resolution (ranging from ${\sim}1$~kpc to ${\sim}50$~pc scales) in an unprecedented large sample of $70$ nearby galaxies, providing data for $675,000$ lines of sight. For this purpose we compile publicly available observations of the atomic gas and tracers of the recent star formation rate, supplemented by a literature compilation of the metallicity, metallicity gradient, and stellar mass. We trace the molecular gas by inverting the Kennicutt--Schmidt relation which enables us to bypass the difficulty of detecting CO at low metallicity. Our main interest is in the observed \hi\ columns of star-forming, molecular regions and we study how the \hi\ columns saturate in increasingly gas-rich regions as a function of gas phase metallicity.

Our paper is structured as follows: In \S~\ref{sec:data} we discuss the literature data we have assembled for \hi, SFR, metallicity, and stellar mass. We discuss how we rescale the metallicities using the SDSS mass--metallicity relation, and describe our methodology for extracting  the \hi\ surface densities as functions of total gas column and metallicity. In \S~\ref{sec:results} we present our main result, that the maximal \hi\ surface density varies inversely with metallicity over a large dynamic range. In \S~\ref{sec:discussion} we discuss our results in the context of shielding-based theoretical models. We summarize our results in \S~\ref{sec:summary}.

\section{Data}
\label{sec:data}


\begin{figure*}[htb]
\epsscale{1.15}
\plotone{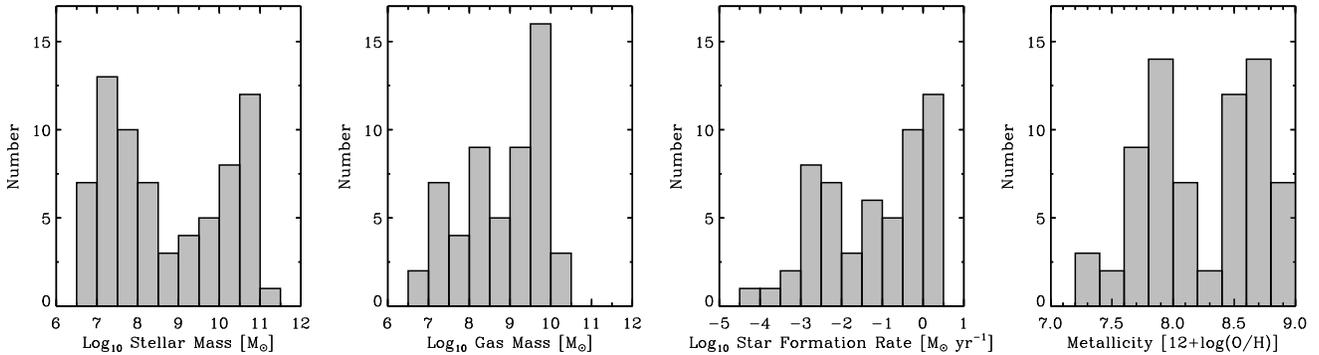}
\vspace*{-2ex} 
\caption{\edit1{Global properties of our galaxy sample showing stellar mass, total gas mass, star formation rate, and gas phase metallicity (from left to right). Metallicities have been rescaled to match the MZR by \citet[][see text]{Andrews13}.}
\label{fig:sample}}
\end{figure*}

For our study, we compile \hi\ and SFR surface density maps, using $21$~cm observations (\S~\ref{sec:atomicgas}) and FUV and $24~\micron$ photometry (\S~\ref{sec:sftracers}). As we discuss further below (\S~\ref{sec:molgas}) we use the SFR to estimate the H$_2$ mass by inverting the Kennicutt--Schmidt relation. We have been able to collect such data for a sample of $70$ nearby galaxies. We supplement these with estimates of the gas phase metallicity, metallicity gradient, stellar mass, distance, inclination, and position angle, all drawn from the literature. \edit1{Figure~\ref{fig:sample} shows the global properties of our galaxy sample,} and Table~\ref{tab:sample} lists these parameters with their literature references. We omit inclinations and position angles here, this information can be found in LEDA \citep{Makarov14}. Our sample is restricted to galaxies that have \hi\ data with linear resolutions better than $1$~kpc, inclinations of less than $76$ degrees, and which are not dominated by a (central) starburst. \edit1{The exploited \hi\ surveys preferentially target star-forming galaxies detected in the infrared but lack well defined selection criteria.} We also require consistent metallicity measurements as we discuss below (\S~\ref{sec:metal}).

\subsection{Atomic Gas}
\label{sec:atomicgas}

The \hi\ data are taken from the VLA surveys THINGS \citep{Walter08}, LITTLE THINGS \citep{Hunter12}, VLA-ANGST \citep{Ott12}, and a recent extension of THINGS for spiral galaxies also surveyed by \emph{Herschel} (NGC 2798, 3049, 3190, 3938, 4236, 4321, 4536, 4594, 4625, 4725, 5474; to be presented by Leroy, Sandstrom, Schruba et al.). The angular resolution is between $5.8 {-} 24.8$ arcsec with a median of $11.7$~arcsec. The linear resolutions correspond to $31 {-} 1085$~pc with a median of $275$~pc. We include the Local Group dwarf galaxy SagDIG with resolution of $31$~arcsec $\approx 166$~pc. The spectral resolution and sensitivity of all \hi\ data sets are more than sufficient to construct integrated intensity maps to radii much larger than considered here. We use the standard optically thin conversion from velocity integrated \hi\ surface brightness, $T_{\rm mb} \Delta v$, to inclination corrected mass surface density, $\Sigma_\ho\, [\Msunperpc] = 0.01986\, T_{\rm mb} \Delta v\, [\Kkmpers]\, \cos(i)$, which includes a factor $1.36$ to account for heavy elements \citep[e.g.,][]{Walter08}.

\subsection{Star Formation Tracers}
\label{sec:sftracers}

We derive star formation rates (SFR) from the combination of \emph{GALEX} FUV ($1350 {-} 1750$~\AA) and \emph{Spitzer} $24$~\micron\ maps. The FUV data are taken by the \emph{GALEX} Nearby Galaxy Survey \citep[NGS;][]{GildePaz07} and the $24$~\micron\ data from the \emph{Spitzer} SINGS and LVL surveys \citep{Dale09a, Dale09b}. We mask foreground stars via their UV color following \citet{Leroy12} and subtract a constant background where necessary. The SFRs are derived as $\Sigma_{\rm SFR}\, [\Msunperyrperkpc] = \left( 8.1{\times}10^{-2}\, I_{\rm FUV} + 3.2{\times}10^{-3}\, I_{\rm 24\, \mu m} \right)\, [\MJypersr]\, \cos(i)$, with the calibration by \citet[][their Appendix~D]{Leroy08}.

\subsection{Molecular Gas}
\label{sec:molgas}

We estimate molecular gas masses by inverting the Kennicutt--Schmidt relation, such that 
\begin{equation}
\label{eq:molgas}
\Sigma_{\rm H_2} \ = \ \Sigma_{\rm SFR} \times \tau_{\rm dep} \ ,
\end{equation}
where $\Sigma_{\rm H_2}$ and $\Sigma_{\rm SFR}$ are the molecular gas mass and SFR surface densities, and $\tau_{\rm dep}$ is the molecular depletion time. We adopt a constant depletion time $\tau_{\rm dep} = 2$~Gyr for all galaxies \citep[e.g.,][]{Schruba12}. This value has been measured for massive galaxies with metallicities above half solar using CO observations \citep[with significant overlap in the galaxy sample used here;][]{Bigiel08, Bigiel11, Schruba11, Leroy13b}. The depletion time for low mass, low metallicity galaxies is less well known, and recent studies have either inferred shorter \citep[e.g.,][]{Hunt15, Amorin16, Grossi16, Jameson16} or longer depletion times \citep{Filho16}. We therefore adopt the same $\tau_{\rm dep}=2$~Gyr for the low mass galaxies. Our use of the Kennicutt--Schmidt relation enables us to identify lines of sight with sufficient molecular columns to have undergone an \hi-to-H$_2$ transition. As our main focus is on the maximal \hi\ surface density and its dependence on metallicity, our results are weakly sensitive to the value of $\tau_{\rm dep}$.

\subsection{Metallicities}
\label{sec:metal}

It is well known that gas phase metallicities\footnote{Following the convention in the field, we use the term `metallicity' as synonym for measurements of the gas phase `oxygen abundance'.} derived from different optical lines and using different calibrations can be subject to large systematic offsets and limitations in their dynamic range \citep{Kewley08}. Here we attempt to compile metallicities and metallicity gradients from the literature with a high degree of internal consistency. We prefer metallicities derived with the direct, electron temperature based method \citep{Pagel92, Izotov06} \edit1{as available for half of our sample}. Where not available (i.e., for most massive galaxies) we adopt measurements that use the $R_{23}$ strong line method using the PT05 calibration \citep{Pilyugin05} that accounts for metallicities derived using the direct method. For detailed discussion of metallicity calibrations we refer to \citet{Kewley08, Moustakas10, Andrews13}.

\begin{figure*}[htb]
\epsscale{1.15}
\plottwo{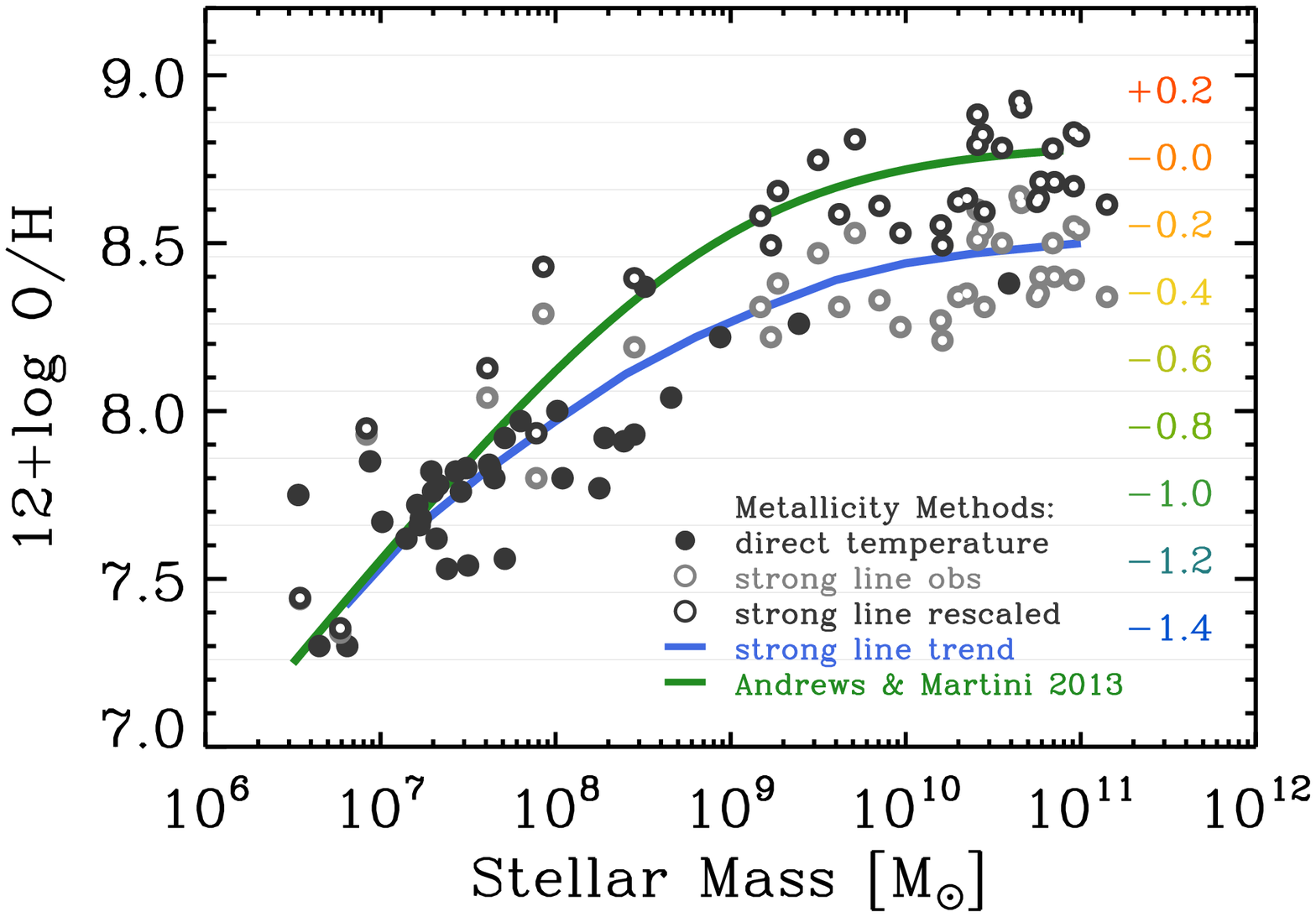}{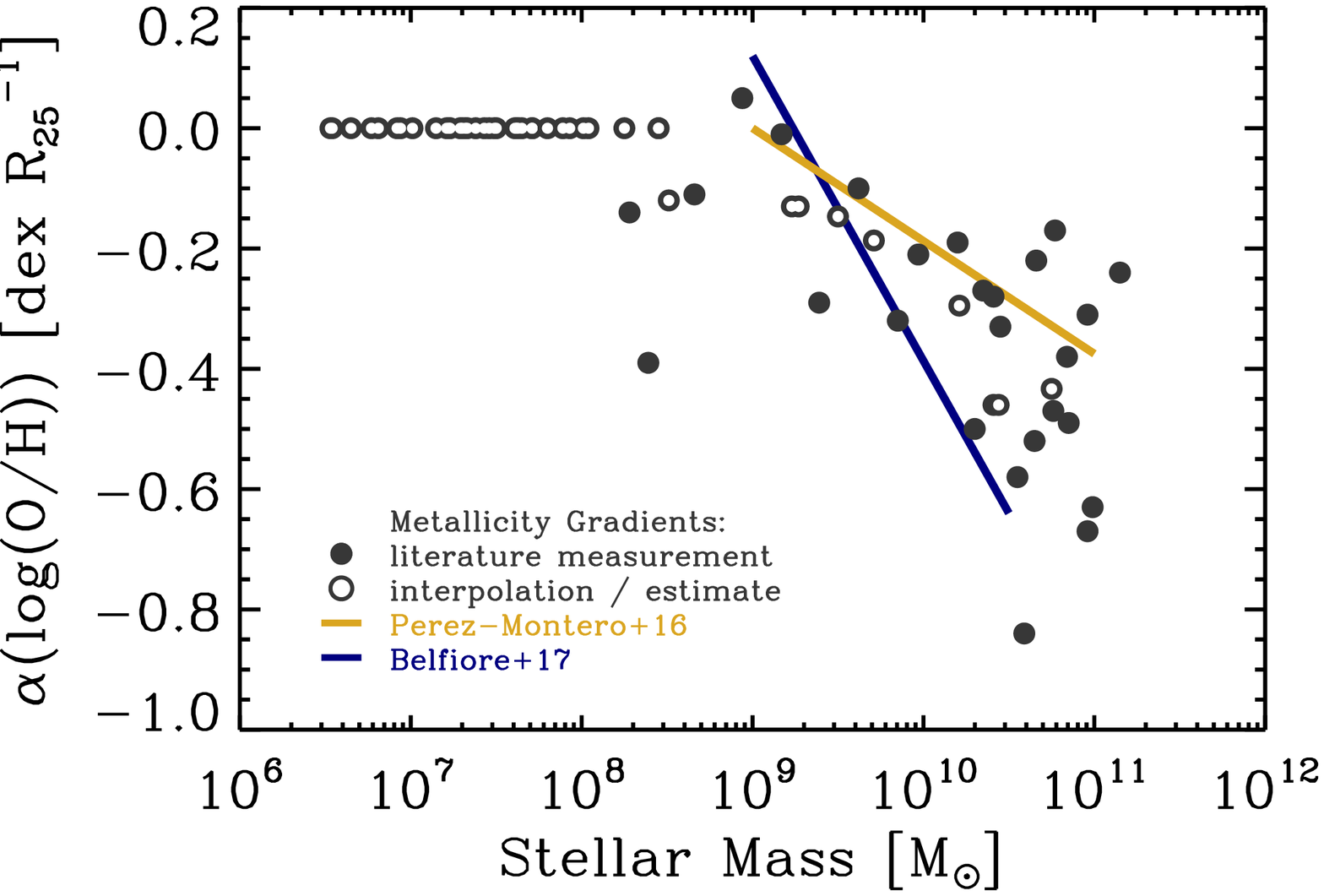}
\vspace*{-2ex} 
\caption{Diagnostic plots of the mass--metallicity (MZR; left) and mass--metallicity gradient (right) relationships. The metallicities are taken from the literature and have been derived with the direct electron temperature or strong line method (solid or open circles, respectively); their median trend is shown by the blue line. We rescale the \edit1{strong line metallicities} (dark open circles) to overlap the MZR derived with the direct electron temperature method for stacked SDSS galaxy spectra by \citet[][green line]{Andrews13} while preserving the observed scatter. \edit1{Colored numbers mark the location of adopted metallicity bins in units of $\log_{10} Z^\prime$.} We complement these by measurements of the metallicity gradient from the literature (filled circles) and where not available we estimate the metallicity gradients (open circles; see text).
\label{fig:metal}}
\end{figure*}

The left panel of Figure~\ref{fig:metal} shows our compiled metallicities versus galaxy stellar mass (grey points). These data are also listed in Table~\ref{tab:sample} together with their literature references. For massive spiral galaxies \edit1{for which measurements of metallicity gradients are available,} we adopt the `characteristic' metallicity measured at galactocentric radii of $0.4~R_{25}$ \edit1{as the galaxy's `average' metallicity} (\citealp{Moustakas10}; which is roughly two times the effective radius\footnote{As derived for the CALIFA galaxy sample with $M_\star \approx 10^9 {-} 10^{11}~\Msun$.}; \citealp{PerezMontero16}). For low mass galaxies we adopt \edit1{measurements averaged over the entire galaxy}.

Our compilation suffers from the known fact that for high mass and high metallicity galaxies the $R_{23}$ strong line method and the PT05 calibration systematically underestimate the true metallicities  by a factor of two \citep[e.g.,][]{Kewley08, Moustakas10}. To remedy this bias, we employ the mass--metallicity relationship (MZR) given by \citet{Andrews13} to correct for the true location of \edit1{strong line} metallicities. \citeauthor{Andrews13} stacked a large sample of SDSS spectra to obtain metallicities using the direct, electron temperature method for a large range of stellar mass: $3{\times}10^{7} - 3{\times}10^{10}~\Msun$. \edit1{They fitted their MZR with an asymptotic logarithmic formula \citep[suggested by][]{Moustakas11}} which is shown as the green line in the left panel of Figure~\ref{fig:metal}. The \edit1{blue curve} shows the median trend for \edit1{our compiled strong line metallicities (light grey open circles)}. We rescale these \edit1{strong line} metallicities by shifting them upwards by the difference between the two curves\footnote{\edit1{The MZR by \citet{Andrews13} does not extend to the lowest and highest stellar masses in our sample. For low mass galaxies differences between direct and strong line metallicities are insignificant and no rescaling is required. For massive galaxies the MZR approaches an asymptotic value of $12 + {\rm log (O/H)} = 8.8$.}}. These rescaled metallicities are shown by the \edit1{dark grey open circles}, and their shifts are listed in the last column of Table~\ref{tab:sample}. With this procedure our sample is then consistent with the \citeauthor{Andrews13} mass--metallicity relation \edit1{(obtained with a single metallicity method), while retaining the scatter (${\sim}0.15$~dex) in our original compilation. This scatter is similar to the scatter of SFR-binned data (${\sim}0.05{-}0.10$~dex) in the MZR of \citet[][their Fig.~11]{Andrews13}, though their measurement (as it inherently is determined from stacked spectra) is a lower limit to the scatter among individual galaxies measured for our sample. The method of metallicity rescaling does not qualitatively affect our results and main conclusions (see \S~\ref{sec:saturation}~\&~\ref{sec:saturation_vs_Z}).}

The right panel of Figure~\ref{fig:metal} shows radial metallicity gradients versus galaxy stellar mass. \edit1{For 27 (massive) galaxies} we were able to collect measurements of the metallicity gradients (filled points). \edit1{For 6 massive galaxies} we estimate the gradient by interpolating the measured gradients of the four galaxies closest in stellar mass (open points) and assign a conservative uncertainty of $0.2$~dex, which equals the spread in measured metallicity gradients at high stellar mass. \edit1{For the remaining 37} low mass galaxies ($M_\star \lesssim 3 {\times} 10^8~\Msun$) we adopt a constant metallicity which is consistent with the finding of negligible gradients in late type, low mass galaxies \citep{Pilyugin14}. We find consistency between our metallicity gradients and determinations by the CALIFA and SAMI surveys \citep[yellow and blue lines;][]{PerezMontero16, Belfiore17}.

To compare observations and theoretical models we consider the so called `Galactic concordance abundance' measured for $29$ main sequence B~stars in the solar neighborhood \citep[$12 + {\rm log (O/H)} = 8.76$;][]{Nieva12, Nicholls17}. For a Galactic metallicity gradient of $\alpha({\rm log(O/H)}) \approx {-}0.04 \ldots {-}0.06$ dex~kpc$^{-1}$ \citep[][and references therein]{Esteban17, Belfiore17}, this translates to $12 {+} {\rm log (O/H)} \approx 8.91$ at $R_{\rm gal} = 0.4 R_{25} \approx 4.6~{\rm kpc}$ \citep{deVaucouleurs78} \edit1{as the `characteristic' metallicity of the Milky Way}. This value is a factor ${\sim}1.4$ above the metallicity of $12 {+} {\rm log (O/H)} = 8.77$ of the \citeauthor{Andrews13} MZR at stellar mass of the Milky Way \citep[$M_\star \approx 6 {\times} 10^{10}~\Msun$;][]{Bovy13}, \edit1{but remains within the ${\sim}1{-}2 \sigma$ scatter of the MZR at $\mathit{SFR} \approx 1$~\Msunperyr\ (see above and their Fig.~11)}. We therefore conclude that the Milky Way metallicity and its gradient are consistent with the rescaled metallicities \edit1{(dark grey points)} that we adopt for our galaxy sample. Hereafter, we denote metallicities by $Z^\prime$ which are normalized to the solar neighborhood metallicity of ${\rm 12+log(O/H)} = 8.76$.

\subsection{Methodology}
\label{sec:method}

For our sample of $70$ galaxies we consider a total of $675{,}500$ individual lines of sight (LOS). We extract our measurements at heterogeneous spatial resolution set by the angular resolution of the VLA \hi\ data ($5.8\arcsec {-} 24.8\arcsec$). For each galaxy we convolve the SFR maps to the \hi\ resolution. We sample these maps on a hexagonal grid with half-beam spacing resulting in an oversampling of four. For each LOS we extract the \hi\ and H$_2$ surface densities and the metallicity, all across the disks of our galaxies. We also compute the local metallicity when adopting a metallicity gradient. We divide our sample into two bins of galactocentric radius of $R_{\rm gal} = 0{-}2$ kpc and $0{-}20$ kpc, into nine bins of local metallicity centered at $\log_{10}\,Z^\prime = -1.4, -1.2, \ldots, +0.2$, and in 15 bins of total gas surface density of $\log_{10}\,\Sigma_{\rm \ho+H_2} = 0, 0.2, \ldots, 3$ \Msunperpc\ spaced by $0.2$~dex. We exclude regions with $\Sigma_\ho \le 1$ \Msunperpc\ or $R_{\rm gal} \ge 2 R_{25}$ for which we do not expect any significant molecular gas reservoirs \citep{Schruba11}. For our radial range we basically include the entire SFR for each galaxy, and hence all of the molecular gas mass, by definition. The radial range also includes the majority of the atomic gas mass. 

We bin our sample of individual LOS by galactocentric radius, metallicity, and total gas surface density. For each bin we calculate the mean \hi\ surface density and the uncertainty in the mean. We determine the uncertainty in the mean using $100$ Monte-Carlo realizations of our data that account for (a) measurement uncertainties in metallicities and their gradients (i.e., the spill over of individual galaxies or regions therein into neighboring metallicity bins),\footnote{Uncertainties in the \hi\ data are small and do not affect our results.} and (b) sample variance via bootstrapping. These uncertainties are shown as error bars in Figures \mbox{\ref{fig:saturation}\,--\,\ref{fig:mscale}}. The scatter in \hi\ surface densities among all LOS contributing to a data point (not shown) is ${\sim}0.4$~dex or ${\sim}2$ times the uncertainty in the mean. More importantly, the scatter is a factor ${\sim}3$ smaller than the systematic variation of the \hi\ saturation with metallicity.

We verified that our binned measurements are insensitive to (a) the location of the bins, (b) individual galaxies with large angular extent or high resolution data, (c) heterogeneous versus fixed linear resolution, or (d) when a subset of galaxies with low inclination is chosen. We performed test (b) by down-weighting LOS in large galaxies or high resolution data to contribute equally in radial annuli of $1$~kpc width.

\edit1{To assess potential systematic biases in our results due to the adopted metallicity determination -- we find them to be minor as can be seen comparing the three rows of panels in Figure~\ref{fig:saturation} -- we repeat our analysis for three different determinations: (a) global metallicities derived with either direct or strong line method as compiled from the literature (top row), (b) global metallicities where strong line values have been rescaled to match the MZR by \citet[][middle row]{Andrews13}, and (c) these rescaled metallicities where also metallicity gradients have been accounted for (bottom row). The first approach resembles previous studies with smaller statistics and coarser resolution. It leverages directly observed metallicities for our entire sample but suffers from limitations inherent to the strong line method: a limited dynamic range in metallicities of massive galaxies which inadvertently remain below the Milky Way metallicity. The second approach remedies this limitation by rescaling the strong line metallicities onto the MZR by \citet{Andrews13} and thus achieves a realistic dynamic range for global metallicities, but neglects genuine metallicity gradients. The latter approach also accounts for metallicity gradients which at least for massive spirals are significant ($\alpha (Z^\prime) \approx \mbox{$-$0.2} \ldots \mbox{$-$0.6}$~dex~$R_{25}^{-1}$). However, as described in \S~\ref{sec:metal} and listed in Table~\ref{tab:sample}, measurements of metallicity gradients are only available for a subset ($27/70$) of our galaxies, so that we have to rely on estimates of the gradients for the remaining galaxies. We are rewarded by an increased and more realistic dynamic range for metallicities of individual lines of sight which, in particular, is important for the inner regions of massive galaxies. We deem these latter metallicities to be most realistic for our sample of galaxy-resolved measurements and concentrate our discussion on these determinations. For all three approaches, the \hi\ saturation columns show a clear inverse dependence on metallicity.}

\section{Results}
\label{sec:results}

\subsection{Saturation of the \hi\ Columns}
\label{sec:saturation}

\begin{figure*}[t]
\epsscale{1.15}
\plottwo{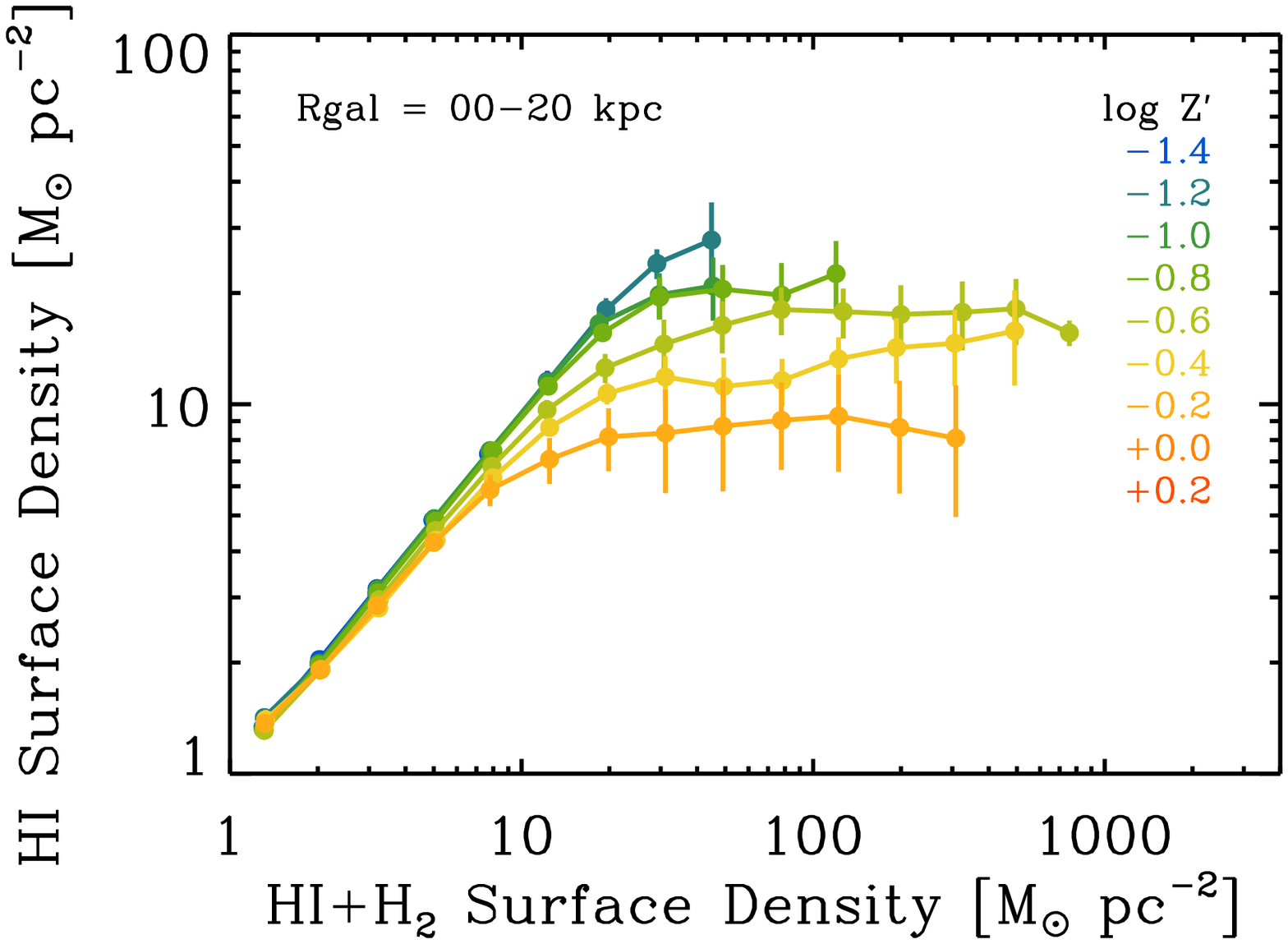}{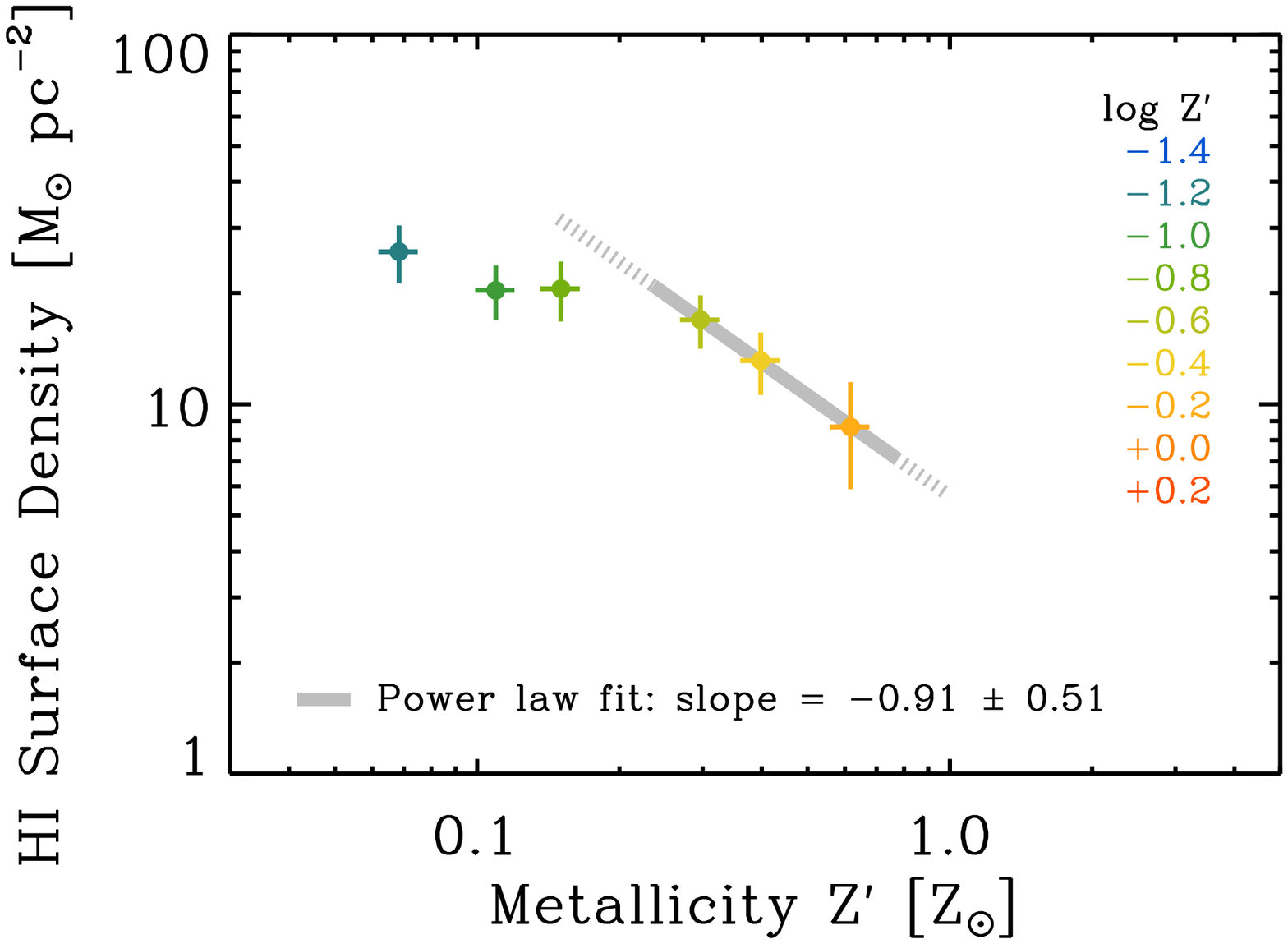}
\plottwo{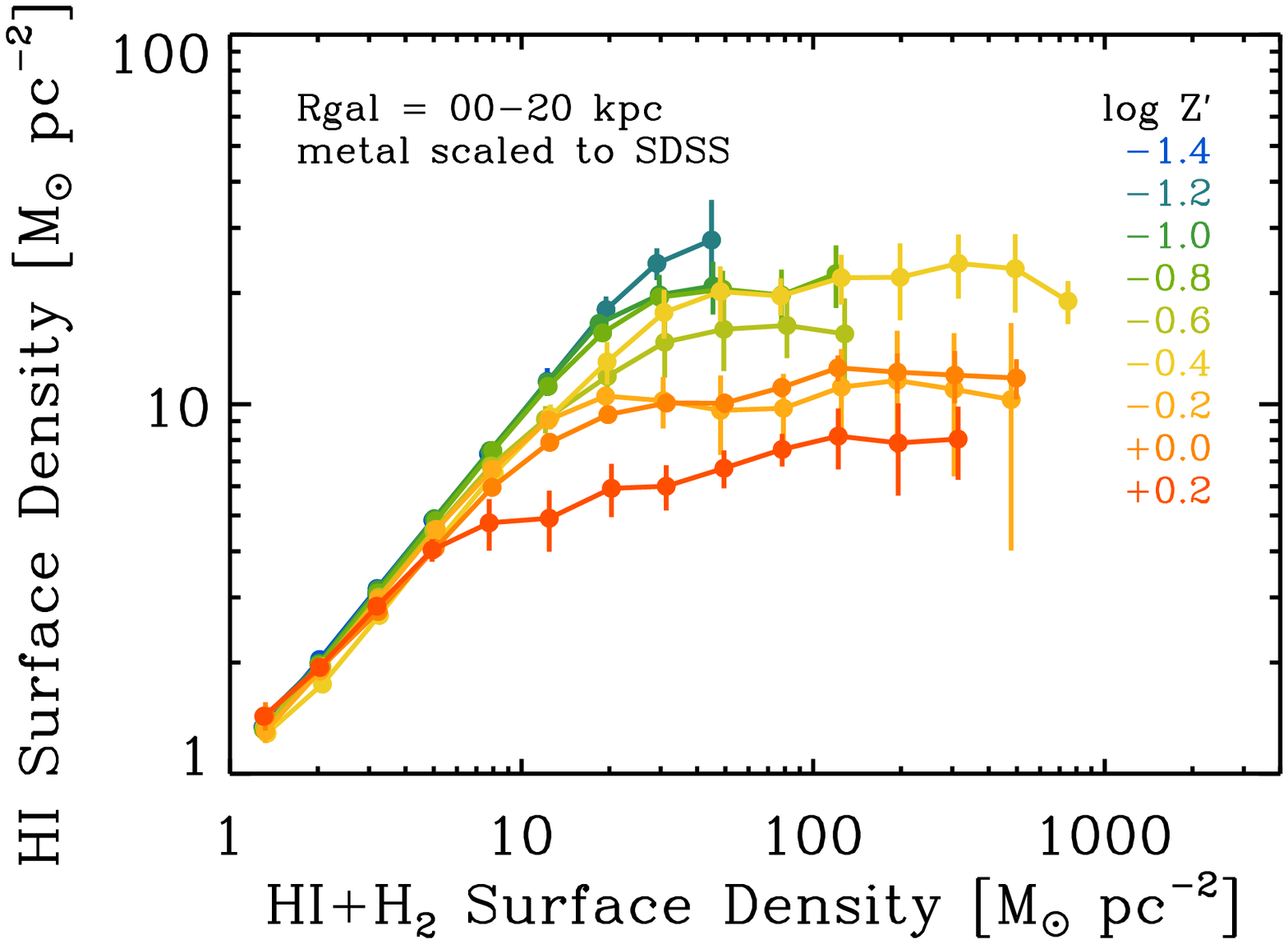}{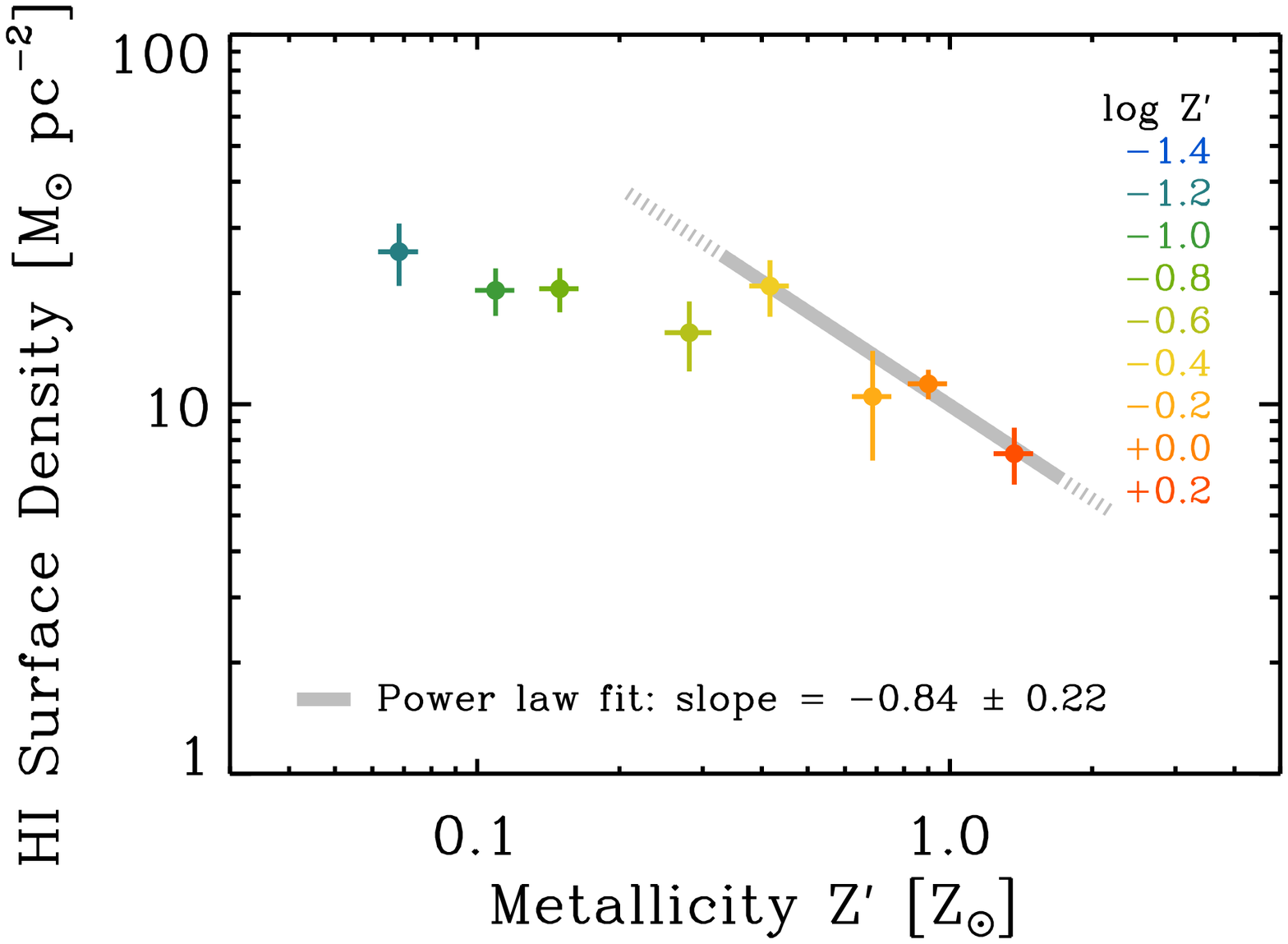}
\plottwo{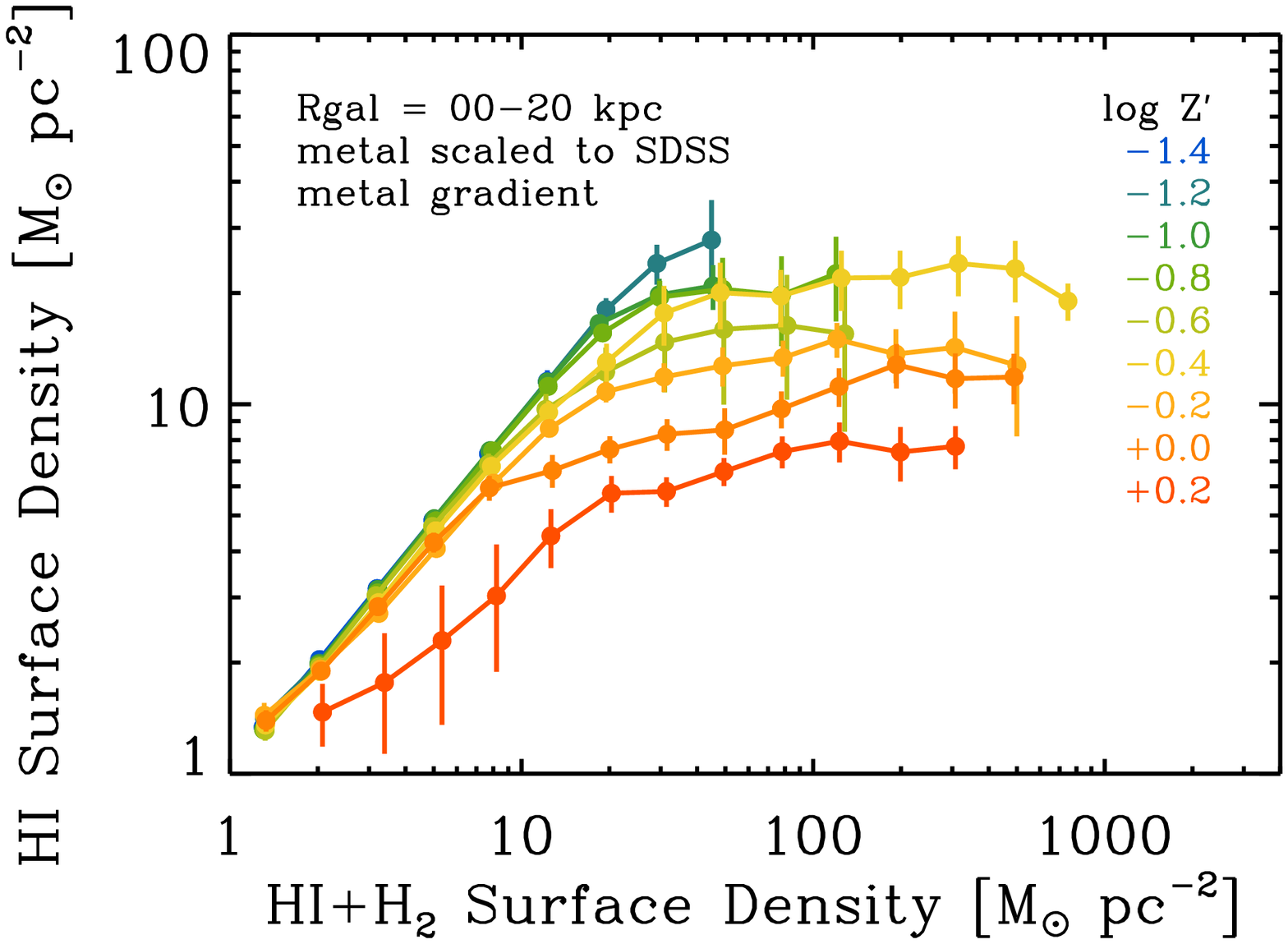}{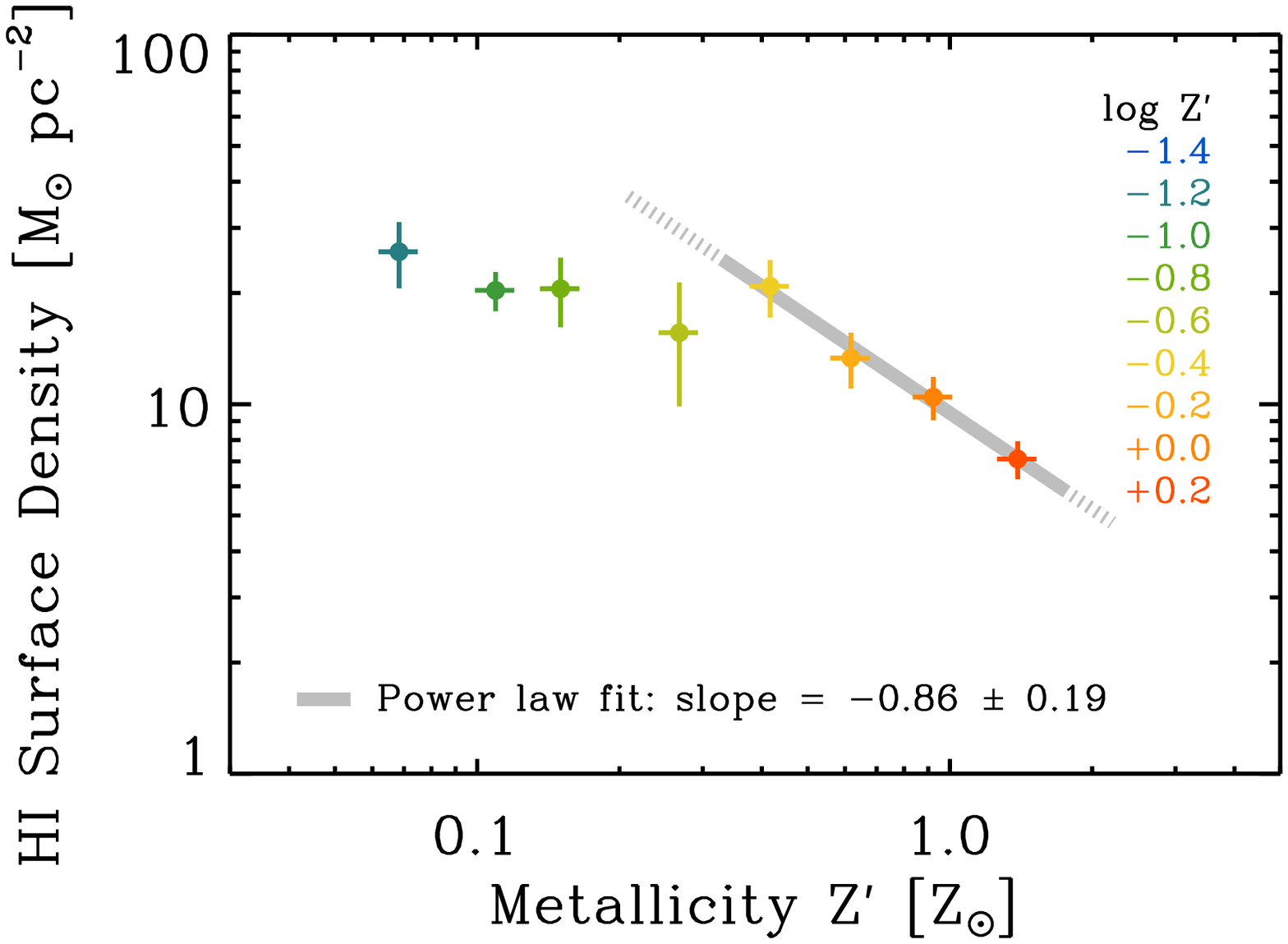}
\vspace*{-2ex} 
\caption{\edit1{\hi\ saturation as function of total gas mass surface density (left) and metallicity (right). We vary the metallicity measurements -- defining into which metallicity bin a line of sight falls -- between global metallicities as compiled from the literature (top), those after strong line metallicities are rescaled to match SDSS direct temperature metallicities (middle), and those where in addition metallicity gradients are applied (bottom). Error bars show uncertainties due to sample variance and uncertainty in the metallicities and their gradients.}
\label{fig:saturation}}
\end{figure*}

\edit1{Figure~\ref{fig:saturation} shows the mean \hi\ surface density as a function of the total gas ($\hi+{\rm H}_2$) surface density (left column) and the maximal the \hi\ surface density as function of metallicity (right column). We vary the metallicity determination between top, middle, and bottom row (see Section~\ref{sec:method}). The left panels show that} at sufficiently small surface densities the H$_2$ contribution is negligible and the total gas surface density equals the \hi\ surface density. As the total gas surface density increases the \hi\ saturates due to \hi-to-H$_2$ conversion, and the H$_2$ finally completely dominates the \hi\ component. For massive spiral galaxies similar to the Milky Way ($M_\star \sim 6 {\times} 10^{10}$ \Msun\ and $Z^\prime \sim 1$), the \hi\ saturates at $\Sigma_\ho \approx 10$~\Msunperpc\ consistent with previous findings \citep[e.g.,][]{Kennicutt98b, Martin01, Wong02, Bigiel08, Schruba11, Wong13, Yim16}.

\edit1{The trend in the \hi\ saturation curves with metallicity is only weakly affected by the adopted metallicity determination (comparing the panels in top, middle, and bottom rows in Figure~\ref{fig:saturation}). This is because the different metallicity determinations change the metallicity value of (massive) galaxies in a very uniform way but do not change the ordering as can be seen by the constant offset between the colored curves in Figures \ref{fig:metal}\,\&\,\ref{fig:saturation}. This changes the value of the metallicities, $Z^\prime$, as referenced to the solar neighborhood metallicity but does not have a significant effect on the trend in the \hi\ saturation as the metallicity changes (i.e., the differential in $Z^\prime$) which is our main result.}

Moreover, we recapitulate that our results are insensitive to our methodology to estimate H$_2$ masses by inverting the Kennicutt--Schmidt relation (Eq.~\ref{eq:molgas}) and the adopted constant molecular gas depletion time of $\tau_{\rm dep} = 2$~Gyr. If some metallicity regimes had a different $\tau_{\rm dep}$ value, then the horizontal extent of the \hi\ saturation curves would be stretched or compressed, but importantly, the measured \hi\ saturation value (i.e., vertical location) remains unaffected.

\subsection{\hi\ Saturation Values as Function of $Z^\prime$}
\label{sec:saturation_vs_Z}

\edit1{In the right panels of Figure~\ref{fig:saturation} we plot the observed maximal \hi\ surface density as a function of metallicity for the three metallicity determinations described above (from top to bottom)}. The \hi\ saturation column varies inversely with metallicity. This is the key result of our paper. Each data point shows the mean \hi\ surface density of all LOS with $\Sigma_{\rm \ho+H_2} \geq 25$ \Msunperpc\ within the respective metallicity bin. As before, error bars show the uncertainty in the mean. We robustly quantify the metallicity dependence with our large sample that probes the \hi\ surface densities from galactic (${\sim}1$~kpc) scales down to cloud (${\sim}50$~pc) scales, and for a wide range of metallicities ($Z^\prime \approx 0.05-2$).

For the $Z^\prime \approx 0.3-2$ range, $\Sigma_{\ho}$ increases almost linearly with decreasing metallicity. Fitting a power law relationship\footnote{We use the IDL routine MPFITEXY that accounts for errors in both variables \citep{Markwardt09, Williams10}} for this range \edit1{to the data points using the rescaled metallicities including gradients (Figure~\ref{fig:saturation}, bottom row)}, we obtain
\begin{equation}
\label{eq:hi_saturation_fit}
\edit1{\log_{10}\,\Sigma_\ho = (-0.86 \pm 0.19)\ \log_{10}\,Z^\prime + (0.98 \pm 0.04)}
\end{equation}
where $\Sigma_\ho$ has units of \Msunperpc\ and $Z^\prime = 1$ for the solar neighborhood metallicity. \edit1{We test the robustness of this result by comparing the three metallicity determinations (top to bottom panels) as well as considering subsamples of galaxies that share metallicities determined with the same method and have observed metallicity gradients. We find their \hi\ saturation curves to be qualitatively the same and their dependence on metallicity to agree within the uncertainties.} Our finding that the \hi\ surface densities increase with decreasing metallicities is consistent with studies of the Magellanic Clouds \citep{Wong09, Bolatto11, Welty12, RomanDuval14} and small samples of nearby dwarf galaxies \citep{Bigiel10b, Fumagalli10, Roychowdhury11, Cormier14, Filho16}. All these studies detected \hi\ surface densities in excess of the canonical $\Sigma_\ho \approx 10$ \Msunperpc\ found for massive, metal rich spiral galaxies \citep{Bigiel08, Schruba11}.

For $Z^\prime \lesssim 0.3$ the \hi\ surface densities \edit1{seem to flatten at} $\Sigma_\ho \approx 20{-}30$ \Msunperpc. However, we note that for low metallicities, the saturation of the \hi\ curves with $\Sigma_{\rm \ho+H_2}$, as seen in \edit1{the left panels of} Fig.~\ref{fig:saturation}), becomes less prominent. The measured $\Sigma_\ho$ values are becoming more uncertain, and may represent lower limits to the true \hi\ saturation values. This results from a lack of sight lines with large total gas surface densities, given our typical measurement scale of $200$~pc \citep{Jameson16, Schruba17a}.

\begin{figure}[t]
\epsscale{1.15}
\plotone{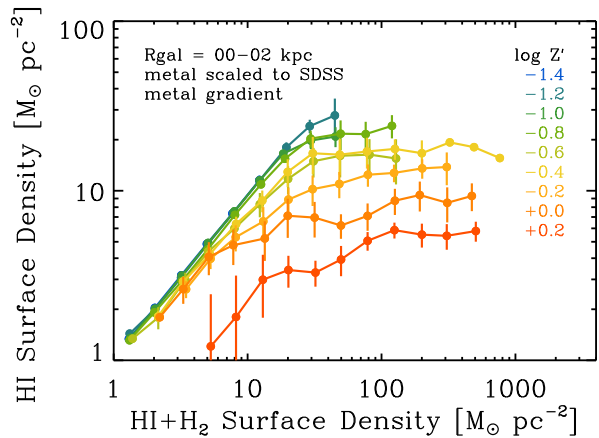}
\vspace*{-2ex} 
\caption{Same as \edit1{bottom left} panel of Figure~\ref{fig:saturation} but selectively for central galaxy at $R_{\rm gal} \leq 2$~kpc where the \hi\ surface densities experience a depression in massive spiral galaxies as compared to their outer disks and lower mass galaxies.
\label{fig:centers}}
\end{figure}

Figure~\ref{fig:centers} shows results for LOS restricted to the inner $2$~kpc of the galaxies and accounting for the metallicity gradients. The maximal \hi\ surface densities are smaller at the highest metallicities compared to \edit1{the bottom left panel of} Figure~\ref{fig:saturation} \citep[see also][]{Bigiel12}. This indicates the possible role of additional parameters affecting the \hi\ saturation at fixed metallicity, such as gas density and UV radiation field.

\begin{figure}[t]
\epsscale{1.15}
\plotone{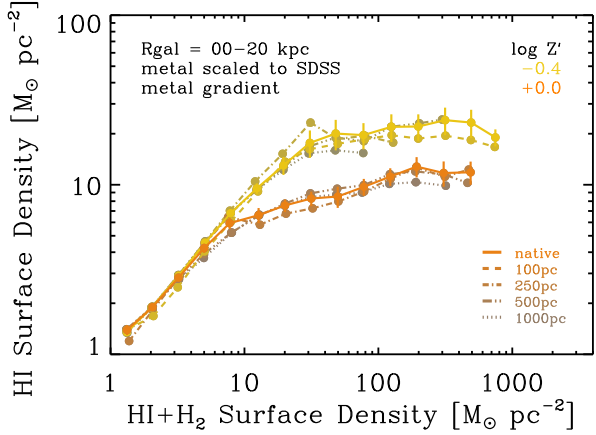}
\vspace*{-2ex} 
\caption{Diagnostic figure highlighting that the \hi\ saturations measured at heterogeneous resolutions (solid lines) for the entire galaxy sample (for two representative metallicity bins) differ insignificantly when determined at various fixed spatial scales (dash-dotted lines) for subsamples which meet the respective resolutions.
\label{fig:mscale}}
\end{figure}

Figure~\ref{fig:mscale} explores the impact of measuring \hi\ surface densities at heterogeneous spatial resolution set by the VLA \hi\ data sets ($5.8\arcsec {-} 24.8\arcsec$ or $31 {-} 1085$~pc). For two representative metallicity bins (at $Z^\prime = 0.4$ and $1$), we convolve the (eligible subsamples of) data to fixed spatial resolutions of $100$, $250$, $500$, $1000$~pc (the dash-dotted lines of differed color saturation). We find that the \hi\ saturation values are only weakly affected by spatial resolution -- reflecting the limited range of volume densities of atomic gas and their high volume filling factor in galaxy disks \citep{Heiles03, Leroy13a} -- and that their variations are of comparable (small) magnitude as the uncertainties in the mean \hi\ saturation values due to sample variance and metallicty (gradient) uncertainties (the error bars in Figures \ref{fig:saturation}\,--\,\ref{fig:centers}). Our results are therefore not biased by the heterogeneous linear resolution at which we perform our measurements.

\section{Discussion}
\label{sec:discussion}

The measurements presented in \S~\ref{sec:results} show that with increasing total gas columns, the \hi\ column saturates, with maximal values scaling inversely with metallicity. Regions with active star formation (treated here to select regions with molecular gas) are also those regions where the \hi\ columns reach the largest values in a galaxy (at fixed metallicity). The H$_2$-dominated regions lie along the rightmost portions of the \hi\ curves, where $\Sigma_{\rm \ho+H_2} \gtrsim 25$ \Msunperpc. At somewhat lower total gas columns the \hi\ may still be in maximal shielding layers but with a subdominant molecular component. At still lower columns the gas is fully atomic along the sight lines.

The number of regions that are rich in molecular gas and where the \hi\ saturates typically occupy only a small fraction of the galaxy's radial extent considered here ($R_{\rm gal} \leq 2~R_{25}$). This implies that the significance of measurements of the \hi-to-H$_2$ transition layers, and even more so of the molecular--atomic ratio, $R_{\rm mol}$, is diluted at low ($\gtrsim$\,kpc) spatial resolution and in azimuthal profiles \citep[such coarse measurement scales are typical for many studies of the Kennicutt--Schmidt relationship especially in dwarf galaxies; e.g.,][]{Elmegreen15, Roychowdhury17, Wang17}. The importance of sufficient spatial resolution has also been emphasized by \citet{Fumagalli10}. These authors compare observations of the \hi\ saturation in dwarf galaxies on $\sim$kpc and on ${\sim}100$~pc scales to predictions by models based on UV shielding \citep{Krumholz09a, Krumholz13, Sternberg14} or midplane pressure \citep{Wong02, Blitz06}. While on kpc-scales both classes of models agree well with the observations, only the shielding-based models are consistent with the observations on $100$~pc scales.

\subsection{Shielding-based Theories}
\label{sec:theory}

Here we discuss predictions of the shielding-based analytic theory by \citet[][hereafter \citetalias{Sternberg14}]{Sternberg14}, we generalize the theory to account for the effects of beam dilution and diffuse \hi\ gas that does not participate in shielding, and we compare the predictions to our observations.

As discussed by \citetalias{Sternberg14}, for a {\it single} optically thick uniform-density two-sided slab irradiated on both sides by an isotropic far-UV flux, the total \hi\ shielding column density is given by
\begin{equation}
\label{eq: HI S14}
N_{\rm \ho,sh} = \frac{1.6}{\sg} \ln \left( \frac{\aG}{3.2} + 1 \right)~.
\end{equation}
In this expression, $\sg \approx 1.9 \times 10^{-21} \ Z^\prime$~cm$^2$ is the dust-absorption cross section over the Lyman--Werner photodissociation band, and $\alpha G$ is the dimensionless ratio of the line-shielded dissociation rate to the H$_2$ formation rate. For a cold neutral medium (CNM) that is in thermal and pressure balance with a warm neutral medium (WNM), $\aG \approx 2.6$ (\citetalias{Sternberg14}, \citealt{Bialy16}). For this typical  $\aG$ value, the \hi\ shielding column~is
\begin{equation}
\label{eq:N_CNM}
N_{\rm \ho,sh} = \frac{5.0 \times 10^{20}}{Z^\prime} \ {\rm cm^{-2}}~,
\end{equation}
or in terms of mass surface density,
\begin{equation}
\label{eq:SD_CNM}
 \Sigma _{\rm \ho,sh} = \frac{5.6}{Z^\prime} \ \Msunperpc~,
\end{equation} including the contribution of Helium.

We stress that Eqs.\ (\ref{eq:N_CNM}\,\&\,\ref{eq:SD_CNM}) apply {\it only for single clouds}, whereas on the coarse resolution of extragalactic observations (typically hundreds of parsecs) the ISM should include several molecular clouds, and potentially also diffuse \hi\ gas that does not contribute to shielding. Let $A_{\rm c}$ be the average size of atomic-molecular complexes (i.e., of molecular clouds and their \hi\ shielding envelopes), $n_{\rm c}$ the number density of such complexes in the galaxy, $H$ the galactic scale height of molecular gas, and $N_{\rm \ho,diff}$ the column density of diffuse \hi\ gas that does not participate in shielding. 
Assuming that the telescope beam is large compared to cloud radii and separation length, the observed \hi\ surface density is 
\begin{align}
\label{eq:SD_obs_1}
\langle \Sigma_\ho \rangle_{\rm obs} & \simeq \underbrace{n_{\rm c} H \varphi_{\rm g} A_{\rm c}  }_{f_{\rm A}} \Sigma_{\rm \ho,sh} + \Sigma_{\rm \ho,diff}~.
\end{align}
where $f_{\rm A} \equiv n_{\rm c} H \varphi_{\rm g} A_{\rm c}$ is the {\it total} area filling factor of molecular clouds (with their \hi\ envelopes) in the telescope beam, and $\varphi_{\rm g}$ is a geometrical factor of order unity ($\varphi_{\rm g} = 1~{\rm or}~2$ for two-sided slabs or thin spherical \hi\ shells, respectively).
Substituting $\Sigma_{\ho, \rm sh}$ from Eq.~(\ref{eq:SD_CNM}) we get
\begin{align}
\label{eq:SD_obs}
\langle \Sigma_{\ho} \rangle_{\rm obs} &= \ \frac{5.6 }{Z^\prime} f_{\rm A} \ \ \Msunperpc + \Sigma_{\rm \ho, diff}~, \\
& = \ \frac{5.6 }{Z^\prime} f_{\rm A} (1+\phi_{\rm diff}) \ \ \Msunperpc \ ,
\label{eq:SD_obs_phi}
\end{align} 
where $\phi_{\rm diff} \equiv \Sigma_{\rm \ho, diff}/(f_{\rm A}\Sigma_{\rm \ho, sh})$ is the total mass residing in the diffuse non-shielding phase relative to the total mass in the shielding envelopes of molecular clouds.

An inverse $1/Z^\prime$ trend for the maximal \hi\ column is naturally produced if $f_{\rm A}$ and $\phi_{\rm diff}$ are independent of metallicity. This is in agreement with the observations within the uncertainty \edit1{(see the right-hand panels of Figure~\ref{fig:saturation}).} If the CNM and WNM phases are associated with the \hi\ shielding envelopes and the diffuse gas respectively\footnote{As the local H$_2$ fraction is ${\propto}\,n$, low density gas is inefficient in (self-) shielding, and the CNM is often considered to dominate the shielding \citep[e.g.,][see however \citealt{Bialy15b} who find that lower density gas may be important for shielding in the Perseus molecular cloud.]{Krumholz09a}}, then the roughly unity  WNM-to-CNM ratio observed in the solar neighborhood \citep{Heiles03} constrains $\phi_{\rm diff} \approx 1$ and consequently (with Eqs.~\ref{eq:hi_saturation_fit}\,\&\,\ref{eq:SD_obs_phi}), $f_{\rm A} \approx 1$. That is, molecular clouds (with their \hi\ envelopes) have an area covering fraction of order unity over the metallicity range, $Z^\prime \approx 0.3-2$. The $1/Z^\prime$ trend may also be obtained for non-constant $f_{\rm A}$ and $\phi_{\rm diff}$. However, this seems unlikely as it requires $f_{\rm A}$ and $\phi_{\rm diff}$ to be fine-tuned such that the product $f_{\rm A} (1+\phi_{\rm diff})$ remains independent of metallicity.

The observed flattening of $\Sigma_\ho$ at low $Z^\prime \lesssim 0.3$ (seen in Figure~\ref{fig:saturation}) suggests that at low metallicities, $f_{\rm A}$ decreases with decreasing metallicity ($\phi_{\rm diff}$ alone cannot explain this trend as the term $(1+\phi_{\rm diff})$ in Eq.~\ref{eq:SD_obs_phi} must be $\geq 1$). This may indicate that molecular clouds are becoming less frequent (smaller $n_{\rm c}$ in Eq.~\ref{eq:SD_obs_1}) or having smaller areas ($A_{\rm c}$) as the metallicity decreases below $Z^\prime \lesssim 0.3$.

\subsection{Observational Constraints on the \hi\ Saturation}

We face the problem that the exact values of $f_{\rm A}$ and $\phi_{\rm diff}$, and their dependence on spatial scale and metallicity are not well known. Above we state the minimal requirements for $f_{\rm A}$, $\phi_{\rm diff}$, and $\Sigma_\ho$ at cloud-scale to explain the observed trend. Here we note that potentially all parameters may vary with measurement scale and metallicity (at least when contrasting small, metal poor and massive, metal rich galaxies).

\emph{Measurement scale:} As discussed in \S~\ref{sec:saturation_vs_Z}, the \hi\ saturation values at low metallicity are uncertain and may represent lower limits to the true saturation values as the curves in Figure~\ref{fig:saturation} do not show a clear turnover. The only high resolution observations of \hi\ at low metallicity are available for the Small Magellanic Cloud (SMC; $Z^\prime = 0.2$). \citet{RomanDuval14} measure $\Sigma_\ho \approx 70$ \Msunperpc\ at $45$~pc scales. This is a factor $\sim 2{-}3$ larger than our measurement of $\Sigma_\ho \approx 20{-}30$ \Msunperpc\ at ${\sim}200$~pc scales. It is expected that molecular clouds are unresolved at $45$~pc \citep{Muller10, Fukui10, Schruba17a}, which together with $f_{\rm A} < 1$ (see below) implies that our measurements of $\Sigma_\ho$ at $200$~pc in low metallicity galaxies are not primarily driven by $\Sigma_{\rm \ho, sh}$ but are sensitive to $\Sigma_{\rm \ho, diff}$.

\emph{Sizes of atomic-molecular complexes:} The contribution of $\Sigma_{\rm \ho, sh}$ relative to $\Sigma_{\rm \ho, diff}$ to the observed $\Sigma_\ho$ is set by the term $f_{\rm A} (1+\phi_{\rm diff})$ in Eq.~\ref{eq:SD_obs_phi} and is driven by two factors: On the one hand, dwarf galaxies typically have low SFR surface densities \citep[e.g.,][]{Bigiel10b, Schruba12} which for a normal molecular cloud population implies a smaller number density of clouds ($n_{\rm c}$). On the other hand, the cloud mass function of dwarf galaxies is shifted to and truncated at smaller cloud masses as compared to massive galaxies \citep[][but also see \citealt{Fukui10}]{Hughes16} which implies a larger $n_{\rm c}$. These two trends at least partially balance one another, though an overall reduction of $f_{\rm A}$ with deceasing metallicity seems plausible (i.e., the drop in SFR surface density is larger than the shift of the cloud mass function).

\emph{Mass balance of shielding versus diffuse atomic gas:} Observations of the solar neighborhood show that the mass ratio of CNM-to-WNM is about unity \citep{Heiles03}. \citet{Warren12} spectrally decompose \hi\ emission maps of $27$ nearby dwarf galaxies, and find the CNM to contribute ${\sim}20\%$ of the total line-of-sight flux when detected (at ${\sim}200$~pc scales), and to contribute less to the global emission. If the diffuse and shielding phases are related to the WNM and CNM phases, this implies that $\phi_{\rm diff}$ increases and $f_{\rm A}$ decreases with decreasing metallicity.

In summary, the trends in $f_{\rm A}$ and $\phi_{\rm diff}$ with spatial scale and metallicity that are required to explain the observed metallicity dependence in the \hi\ saturation value are in good agreement with observations. However, higher resolution \hi\ observations, independent measurements of the CNM/WNM balance and the size of atomic-molecular complexes, and larger statistics are required to robustly determine the relative contribution of the various effects that control the \hi-to-H$_2$ transition at very low metallicity ($Z^\prime < 0.3$).

\section{Summary}
\label{sec:summary}

We investigate the metallicity dependence of the maximal \hi\ surface densities in star-forming regions of $70$ nearby galaxies at spatial scales of $1$~kpc to $50$~pc. To do this, we compile maps of \hi\ and SFR tracers (FUV and $24~\micron$), as well as measurements of stellar mass, gas phase metallicity, and their gradients from the literature (Table~\ref{tab:sample}). Our galaxies cover a wide range in stellar mass ($M_\star \approx 10^{6} - 10^{11}$~\Msun), atomic gas mass ($M_\ho \approx 10^{6} - 10^{10}$~\Msun), star formation rate ($\mathit{SFR} \approx 5{\times}10^{-5} - 4$~\Msunperyr), and metallicity ($Z^\prime \approx 0.05 - 2$) relative to solar. \edit1{We perform our analysis adopting the literature metallicities (derived with two different methods) as well as rescaled metallicities} that match the mass--metallicity relation determined by \citet{Andrews13} for SDSS galaxies employing metallicities derived with the direct electron temperature method. \edit1{The results agree within the uncertainties.} We bin our sample of $675{,}000$ individual lines of sight by galactocentric radius, local metallicity, and total gas surface density. Our main results are:
\begin{itemize}
\item The \hi\ surface density saturates at large total gas columns. The maximal values are significantly anti-correlated with gas phase metallicity in a roughly linearly inverse relationship. A power law fit over $Z^\prime = 0.3-2$ has \edit1{slope of $-0.86 \pm 0.19$} and intercept of $\Sigma_\ho \approx 10 \pm 1$ \Msunperpc. At $Z^\prime < 0.3$ the \hi\ saturates at $\Sigma_\ho \approx 20{-}30$~\Msunperpc\ on scales of $200$~pc.
\item The maximal \hi\ surface densities show little variation with spatial resolution (over $0.1-1$~kpc scales), galactocentric radius, and between galaxies; with the exception of the central parts ($R_{\rm gal} \lesssim 2$~kpc) of massive spiral galaxies that frequently have a depression of \hi\ gas by a factor ${\sim}2$ as compared to the \hi\ columns in their disks.
\item The ${\sim}1/Z^\prime$ dependence of the \hi\ saturation columns observed at $Z^\prime = 0.3-2$ is in good agreement with predictions of analytic theories of the \hi-to-H$_2$ transition by \citet{Krumholz09a, McKee10, Sternberg14} and \citet{Bialy16} that motivate (dust) shielding from dissociating radiation as a prime requirement for molecule formation and survival.
\item At low metallicity ($Z^\prime < 0.3$) the observed flattening at $\Sigma_\ho \approx 20{-}30$ \Msunperpc\ can be reproduced by the models when the area covering fraction of molecular clouds and their atomic envelopes decrease while the diffuse \hi\ gas may dominate over the \hi\ gas in shielding layers. Observational constraints of these parameters remain sparse but those available are consistent with the required trends.
\end{itemize}

Observational studies of the \hi-to-H$_2$ conversion employing large statistics are (slowly) emerging. In particular observations of high spatial resolution at the scale of molecular clouds are an asset which thus far are only available for galaxies in (the vicinity of) the Local Group. Furthermore, robust measurements of total gas columns are desired, which may be accessible from optical or near infrared stellar extincting mapping. This would allow to search for (second-order) dependencies of the \hi\ saturation values in excess of gas phase metallicity: e.g., radiation field strength and gas volume density as proposed by shielding-based theories, but also midplane pressure, self-gravity on small scales, or the stochasticity of the \hi-to-H$_2$ conversion in (low mass) systems where equilibrium-based models no longer apply. 

\acknowledgments

\edit1{The authors thank the anonymous referee for a careful and constructive report that improved this paper.} AS thanks I-Ting Ho and Rob Yates for helpful discussions about metallicity determinations. This work was supported by DFG/DIP grant STE 1869/2-1 GE 625/17-1.

\facilities{VLA, Spitzer, GALEX}

\bibliographystyle{aasjournal}


{\renewcommand{\arraystretch}{1.1} 
\begin{deluxetable*}{lrlcclclcrl}
\tablecolumns{11}
\tabletypesize{\scriptsize}
\tablecaption{Sample Properties\label{tab:sample}}
\tablehead{\colhead{Galaxy} & \colhead{Dist} & \colhead{Ref.} & \colhead{log(M$_\star$)} & \colhead{Ref.} & \colhead{12+log(O/H)} & \colhead{Type\tablenotemark{a}} & \colhead{Ref.} & \colhead{$\Delta$(O/H)\tablenotemark{b}} & \colhead{$\alpha$(O/H)} & \colhead{Ref.} \\
\colhead{} & \colhead{Mpc} & \colhead{} & \colhead{(\Msun)} & \colhead{} & \colhead{dex} & \colhead{} & \colhead{} & \colhead{dex} & \colhead{(dex R$_{25}^{-1}$)} & \colhead{}}
\startdata
DDO053   & 3.6 & K11 & 7.29 & C14 & 7.82 $\pm$ 0.09 & d & B12 & 0.04 & 0.00 & S18 \\[-1ex]
DDO154   & 3.8 & \nodata & 7.01 & MM15 & 7.67 $\pm$ 0.06 & d & MA10 & 0.02 & 0.00 & S18 \\[-1ex]
HOI      & 3.9 & K11 & 7.71 & C14 & 7.92 $\pm$ 0.05 & d & B12 & 0.11 & 0.00 & S18 \\[-1ex]
HOII     & 3.4 & \nodata & 8.28 & MM15 & 7.92 $\pm$ 0.10 & d & B12 & 0.19 & -0.14 $\pm$ 0.08 & P15 \\[-1ex]
IC2574   & 4.0 & \nodata & 8.45 & MM15 & 7.93 $\pm$ 0.05 & d & B12 & 0.21 & 0.00 & S18 \\[-1ex]
M81DWA   & 3.6 & \nodata & 6.77 & C14 & 7.34 $\pm$ 0.20 & s & MO10 & 0.01 & 0.00 & S18 \\[-1ex]
M81DWB   & 3.6 & \nodata & 7.33 & MM15 & 7.78 $\pm$ 0.05 & d & B12 & 0.05 & 0.00 & S18 \\[-1ex]
NGC0628  & 9.7 & MQ17 & 10.35 & MM15 & 8.35 $\pm$ 0.01 & s & MO10 & 0.28 & -0.27 $\pm$ 0.05 & MO10 \\[-1ex]
NGC0925  & 9.1 & K11 & 9.97 & RR15 & 8.25 $\pm$ 0.01 & s & MO10 & 0.28 & -0.21 $\pm$ 0.03 & MO10 \\[-1ex]
NGC2366  & 3.2 & \nodata & 8.39 & C14 & 7.91 $\pm$ 0.05 & d & B12 & 0.20 & -0.39 $\pm$ 0.06 & P14 \\[-1ex]
NGC2403  & 3.2 & K11 & 9.85 & C14 & 8.33 $\pm$ 0.01 & s & MO10 & 0.28 & -0.32 $\pm$ 0.03 & MO10 \\[-1ex]
NGC2841  & 14.1 & K11 & 10.99 & MM15 & 8.54 $\pm$ 0.03 & s & MO10 & 0.28 & -0.63 $\pm$ 0.46 & MO10 \\[-1ex]
NGC2903  & 8.9 & K11 & 10.65 & MM15 & 8.64 $\pm$ 0.05 & s & E08 & 0.28 & -0.52 $\pm$ 0.07 & P14 \\[-1ex]
NGC2976  & 3.6 & K11 & 9.27 & MM15 & 8.38 $\pm$ 0.06 & s & MO10 & 0.28 & -0.13 $\pm$ 0.20 & S18 \\[-1ex]
NGC3049  & 19.2 & K11 & 9.71 & RR15 & 8.53 $\pm$ 0.01 & s & MO10 & 0.28 & -0.19 $\pm$ 0.20 & S18 \\[-1ex]
NGC3184  & 11.8 & K11 & 10.41 & MM15 & 8.51 $\pm$ 0.01 & s & MO10 & 0.28 & -0.46 $\pm$ 0.06 & MO10 \\[-1ex]
NGC3198  & 14.1 & K11 & 10.30 & MM15 & 8.34 $\pm$ 0.02 & s & MO10 & 0.28 & -0.50 $\pm$ 0.14 & MO10 \\[-1ex]
NGC3351  & 9.3 & K11 & 10.41 & MM15 & 8.60 $\pm$ 0.01 & s & MO10 & 0.28 & -0.28 $\pm$ 0.04 & MO10 \\[-1ex]
NGC3521  & 11.2 & K11 & 10.96 & MM15 & 8.39 $\pm$ 0.02 & s & MO10 & 0.28 & -0.67 $\pm$ 0.10 & P14 \\[-1ex]
NGC3621  & 6.6 & K11 & 10.20 & RR15 & 8.27 $\pm$ 0.02 & s & MO10 & 0.28 & -0.19 $\pm$ 0.13 & MO10 \\[-1ex]
NGC3627  & 9.4 & K11 & 10.75 & MM15 & 8.34 $\pm$ 0.24 & s & MO10 & 0.28 & -0.43 $\pm$ 0.20 & S18 \\[-1ex]
NGC3938  & 17.9 & K11 & 10.55 & MM15 & 8.50 $\pm$ 0.10 & s & P14 & 0.28 & -0.58 $\pm$ 0.09 & P14 \\[-1ex]
NGC4214  & 2.9 & K11 & 8.94 & MM15 & 8.22 $\pm$ 0.05 & d & B12 & 0.26 & 0.05 $\pm$ 0.12 & P15 \\[-1ex]
NGC4236  & 4.4 & K11 & 9.23 & C14 & 8.22 $\pm$ 0.20 & s & RR14 & 0.27 & -0.13 $\pm$ 0.20 & S18 \\[-1ex]
NGC4321  & 14.3 & K11 & 10.84 & MM15 & 8.50 $\pm$ 0.03 & s & MO10 & 0.28 & -0.38 $\pm$ 0.21 & MO10 \\[-1ex]
NGC4449  & 4.2 & K11 & 9.39 & MM15 & 8.26 $\pm$ 0.09 & d & B12 & 0.28 & -0.29 $\pm$ 0.08 & A17 \\[-1ex]
NGC4536  & 14.5 & K11 & 10.21 & MM15 & 8.21 $\pm$ 0.08 & s & MO10 & 0.28 & -0.29 $\pm$ 0.20 & S18 \\[-1ex]
NGC4725  & 11.9 & K11 & 10.76 & MM15 & 8.35 $\pm$ 0.13 & s & MO10 & 0.28 & -0.47 $\pm$ 0.08 & P14 \\[-1ex]
NGC4736  & 4.7 & K11 & 10.45 & MM15 & 8.31 $\pm$ 0.03 & s & MO10 & 0.28 & -0.33 $\pm$ 0.18 & MO10 \\[-1ex]
NGC4826  & 5.3 & K11 & 10.44 & MM15 & 8.54 $\pm$ 0.10 & s & MO10 & 0.28 & -0.46 $\pm$ 0.20 & S18 \\[-1ex]
NGC5055  & 8.9 & MQ17 & 10.85 & MM15 & 8.40 $\pm$ 0.03 & s & MO10 & 0.28 & -0.49 $\pm$ 0.03 & P14 \\[-1ex]
NGC5194  & 8.6 & MQ17 & 10.96 & MM15 & 8.55 $\pm$ 0.01 & s & MO10 & 0.28 & -0.31 $\pm$ 0.06 & MO10 \\[-1ex]
NGC5236  & 4.5 & K11 & 10.66 & MM15 & 8.62 $\pm$ 0.01 & s & E08 & 0.28 & -0.22 $\pm$ 0.03 & P14 \\[-1ex]
NGC5457  & 6.7 & K11 & 10.59 & MM15 & 8.38 $\pm$ 0.10 & d & C16 & 0.28 & -0.84 $\pm$ 0.03 & P14 \\[-1ex]
NGC5474  & 6.8 & K11 & 9.17 & MM15 & 8.31 $\pm$ 0.22 & s & MO10 & 0.27 & -0.01 $\pm$ 0.09 & P14 \\[-1ex]
NGC6946  & 6.8 & K11 & 10.77 & RR15 & 8.40 $\pm$ 0.03 & s & MO10 & 0.28 & -0.17 $\pm$ 0.15 & MO10 \\[-1ex]
NGC7331  & 14.5 & K11 & 11.15 & RR15 & 8.34 $\pm$ 0.02 & s & MO10 & 0.27 & -0.24 $\pm$ 0.35 & MO10 \\[-1ex]
NGC7793  & 3.9 & K11 & 9.62 & MM15 & 8.31 $\pm$ 0.02 & s & MO10 & 0.28 & -0.10 $\pm$ 0.08 & MO10 \\[-1ex]
CVNIDWA  & 3.6 & H12 & 6.81 & C14 & 7.30 $\pm$ 0.04 & d & B12 & 0.01 & 0.00 & S18 \\[-1ex]
DDO043   & 7.8 & H12 & 7.63 & nan & 7.83 $\pm$ 0.03 & m & P16 & 0.09 & 0.00 & S18 \\[-1ex]
DDO046   & 6.1 & H12 & 7.15 & nan & 7.62 $\pm$ 0.09 & m & P16 & 0.02 & 0.00 & S18 \\[-1ex]
DDO047   & 5.2 & H12 & 7.46 & nan & 7.76 $\pm$ 0.03 & m & P16 & 0.07 & 0.00 & S18 \\[-1ex]
DDO069   & 0.8 & H12 & 6.65 & C14 & 7.30 $\pm$ 0.05 & d & B12 & 0.01 & 0.00 & S18 \\[-1ex]
DDO070   & 1.3 & H12 & 7.38 & MM15 & 7.53 $\pm$ 0.05 & d & B12 & 0.05 & 0.00 & S18 \\[-1ex]
DDO075   & 1.3 & H12 & 7.50 & MM15 & 7.54 $\pm$ 0.06 & d & B12 & 0.07 & 0.00 & S18 \\[-1ex]
DDO087   & 7.7 & H12 & 7.62 & MM15 & 7.84 $\pm$ 0.04 & d & C09 & 0.09 & 0.00 & S18 \\[-1ex]
DDO101   & 6.4 & H12 & 8.01 & MM15 & 8.00 $\pm$ 0.10 & d & B12 & 0.15 & 0.00 & S18 \\[-1ex]
DDO126   & 4.9 & H12 & 7.89 & MM15 & 7.80 $\pm$ 0.20 & s & H12 & 0.13 & 0.00 & S18 \\[-1ex]
DDO133   & 3.5 & H12 & 7.61 & MM15 & 8.04 $\pm$ 0.20 & s & MA10 & 0.09 & 0.00 & S18 \\[-1ex]
DDO165   & 4.6 & H12 & 8.04 & MM15 & 7.80 $\pm$ 0.06 & d & B12 & 0.16 & 0.00 & S18 \\[-1ex]
DDO167   & 4.2 & H12 & 7.22 & nan & 7.66 $\pm$ 0.20 & m & H12 & 0.03 & 0.00 & S18 \\[-1ex]
DDO168   & 4.3 & H12 & 7.93 & MM15 & 8.29 $\pm$ 0.10 & s & H12 & 0.14 & 0.00 & S18 \\[-1ex]
DDO187   & 2.2 & H12 & 6.53 & MM15 & 7.75 $\pm$ 0.05 & d & B12 & 0.00 & 0.00 & S18 \\[-1ex]
DDO216   & 1.1 & H12 & 6.92 & MM15 & 7.93 $\pm$ 0.14 & s & H12 & 0.02 & 0.00 & S18 \\[-1ex]
HARO29   & 5.9 & H12 & 7.65 & C14 & 7.80 $\pm$ 0.10 & d & MA10 & 0.09 & 0.00 & S18 \\[-1ex]
HARO36   & 9.3 & H12 & 8.51 & MM15 & 8.37 $\pm$ 0.10 & d & MA10 & 0.21 & -0.12 $\pm$ 0.20 & S18 \\[-1ex]
IC1613   & 0.7 & H12 & 7.32 & C14 & 7.62 $\pm$ 0.05 & d & B12 & 0.04 & 0.00 & S18 \\[-1ex]
MRK178   & 3.9 & H12 & 7.43 & C14 & 7.82 $\pm$ 0.06 & d & B12 & 0.06 & 0.00 & S18 \\[-1ex]
NGC1569  & 3.0 & \nodata & 8.45 & J12 & 8.19 $\pm$ 0.02 & s & E08 & 0.21 & 0.00 & S18 \\[-1ex]
NGC3738  & 4.9 & H12 & 8.66 & MM15 & 8.04 $\pm$ 0.05 & d & B12 & 0.23 & -0.11 $\pm$ 0.10 & P15 \\[-1ex]
NGC4163  & 2.9 & H12 & 7.71 & MM15 & 7.56 $\pm$ 0.14 & d & B12 & 0.11 & 0.00 & S18 \\[-1ex]
SAGDIG   & 1.1 & H12 & 6.54 & MC12 & 7.44 $\pm$ 0.15 & s & H12 & 0.00 & 0.00 & S18 \\[-1ex]
UGC8508  & 2.6 & H12 & 7.30 & MM15 & 7.76 $\pm$ 0.07 & d & B12 & 0.04 & 0.00 & S18 \\[-1ex]
VIIZW403 & 4.3 & H12 & 7.21 & RR15 & 7.72 $\pm$ 0.05 & d & E08 & 0.03 & 0.00 & S18 \\[-1ex]
WLM      & 1.0 & H12 & 7.49 & C14 & 7.83 $\pm$ 0.06 & d & B12 & 0.07 & 0.00 & S18 \\[-1ex]
DDO125   & 2.6 & O12 & 7.80 & MM15 & 7.97 $\pm$ 0.06 & d & B12 & 0.12 & 0.00 & S18 \\[-1ex]
DDO181   & 3.1 & O12 & 6.94 & MM15 & 7.85 $\pm$ 0.04 & d & MA10 & 0.02 & 0.00 & S18 \\[-1ex]
NGC0247  & 3.5 & O12 & 9.50 & MM15 & 8.47 $\pm$ 0.20 & s & D06 & 0.28 & -0.15 $\pm$ 0.20 & S18 \\[-1ex]
NGC3109  & 1.3 & O12 & 8.25 & C14 & 7.77 $\pm$ 0.07 & d & MA10 & 0.18 & 0.00 & S18 \\[-1ex]
NGC3741  & 3.2 & O12 & 7.23 & C14 & 7.68 $\pm$ 0.05 & d & B12 & 0.03 & 0.00 & S18\\
\enddata
\tablenotetext{a}{Metallicity methods: d = direct electron temperature, m = mix, s = strong line}
\tablenotetext{b}{Difference between observed metallicities and mass--metallicity relationship by \citet{Andrews13}. Here only applied for strong line metallicities.}
\tablerefs{A17 = \citet{Annibali17}; B12 = \citet{Berg12}; C09 = \citet{Croxall09}; C14 = \citet{Cook14}; C16 = \citet{Croxall16}; D06 = \citet{Davidge06}; E08 = \citet{Engelbracht08}; H12 = \citet{Hunter12}; J12 = \citet{Johnson12}; K11 = \citet{Kennicutt11}; MA10 = \citet{Marble10}; MC12 = \citet{McConnachie12}; MM15 = \citet{MunozMateos15}; MO10 = \citet{Moustakas10}; MQ17 = \citet{McQuinn17}; O12 = \citet{Ott12}; P14 = \citet{Pilyugin14}; P15 = \citet{Pilyugin15}; P16 = \citet{Pilyugin16}; RR14 = \citet{RemyRuyer14}; RR15 = \citet{RemyRuyer15}; S18 = This paper}
\end{deluxetable*}
} 

\end{document}